\documentclass[submission,copyright,creativecommons]{eptcs}
\providecommand{\event}{ICLP 2021} % Name of the event you are submitting to

\usepackage{underscore}           % Only needed if you use pdflatex.
\usepackage{capt-of}

\title{Fixpoint Semantics for Recursive SHACL}

\author{Bart Bogaerts
\institute{Vrije Universiteit Brussel, Belgium %\thanks{A fine university.}
}
\email{bart.bogaerts@vub.be}
\and
Maxime Jakubowski 
\institute{Universiteit Hasselt, Belgium}
\institute{Vrije Universiteit Brussel, Belgium}
\email{maxime.jakubowski@uhasselt.be}
}
\def\titlerunning{Fixpoint Semantics for Recursive SHACL}
\def\authorrunning{B. Bogaerts \& M. Jakubowski}

\usepackage{booktabs}

\usepackage{paralist} 
\setdefaultenum{\bfseries(i)}{}{}{}  

\newcommand\concept{\m{\phi}}
\newcommand\closed{\m{\mathit{closed}}}

\newcommand{\eq}{\mathit{eq}}
\newcommand{\disj}{\m{\mathit{disj}}}

\newcommand\dom{\m{\Delta}}
\newcommand{\geqn}[3]{\geq_{#1}#2.#3}

\usepackage{stmaryrd}
\newcommand{\iexpr}[2]{\llbracket #1 \rrbracket^{#2}}      % \concept(G), E(G)
\newcommand{\semi}[1]{\iexpr{#1}I}

\usepackage{import}

\usepackage{ifthen}
\usepackage{url}
\usepackage{amssymb}
\usepackage{amsmath}
\usepackage{import}
\usepackage{xparse}
\usepackage{amsmath}
\usepackage[usenames,dvipsnames]{xcolor}
\usepackage{xspace}
\usepackage{datatool}
\usepackage{glossaries}

\providecommand\m[1]{\ensuremath{#1}\xspace}
\renewcommand{\m}[1]{\ensuremath{#1}\xspace}
\newcommand{\trval}[1]{\m{\mathbf{#1}}}

	\newcommand{\lrule}{\leftarrow}
	\newcommand{\cause}{\stackrel{c}{\lrule}}
	
	\newcommand{\ltrue}{\trval{t}}
	\newcommand{\lfalse}{\trval{f}}
	\newcommand{\lunkn}{\trval{u}}
	
	\newcommand{\voc}{\m{\Sigma}}

	\newcommand{\struct}{\m{I}}

	\newcommand{\theory}{\m{\mathcal{T}}}

	\NewDocumentCommand\inter{g+g}{%
	  \IfNoValueTF{#1}
	    {\struct}
	    {\m{#1^{#2}}}}

	\renewcommand{\int}{\m{\mathbb{Z}}}

	\newcommand{\leqp}{\m{\leq_p}}

	\newcommand{\leqt}{\m{\leq_t}}

	\DeclareMathOperator\lfp{lfp}

	\NewDocumentCommand\subs{g+g}{%
	  \IfNoValueTF{#1}
	    {\m{/}}
	    {\m{#1/ #2}}}

	\newcommand{\logicname}[1]{\textsc{#1}\xspace}

	\newcommand{\idp}{\logicname{IDP}}

\newcommand{\ouracronym}[3]{%
	\newacronym{#1}{#2}{#3}
	\expandafter\newcommand\csname #1\endcsname{\gls{#1}\xspace}%
}
	\ouracronym{FO}{FO}{first-order logic}
	\ouracronym{PC}{PC}{propositional calculus}
	\ouracronym{MX}{MX}{Model Expansion}
	\ouracronym{MO}{MO}{Model Optimization}
	\ouracronym{ASP}{ASP}{Answer Set Programming}
	\ouracronym{TP}{TP}{Theorem Proving}
	\ouracronym{LP}{LP}{Logic Programming}
	\ouracronym{CP}{CP}{Constraint Programming}
	\ouracronym{FP}{FP}{Functional Programming}
	\ouracronym{KR}{KR}{Knowledge Representation}
	\ouracronym{CSP}{CSP}{Constraint Satisfaction Problem}
	\ouracronym{SMT}{SMT}{SAT Modulo Theories}
	\ouracronym{KBS}{KBS}{knowledge base system}
	\ouracronym{NNF}{NNF}{Negation Normal Form}
	\ouracronym{FNNF}{FNNF}{Flat Negation Normal Form}
	\ouracronym{DefNNF}{DefNNF}{Definition Negation Normal Form}
	\ouracronym{DEFNF}{DEFNF}{Definition Normal Form}
	\ouracronym{CDCL}{CDCL}{Conflict-Driven Clause-Learning}
	\ouracronym{WFS}{WFS}{Well-Founded Semantics}
	\ouracronym{LCG}{LCG}{Lazy Clause Generation}
	\ouracronym{AEL}{AEL}{Autoepistemic Logic}
	\ouracronym{OEL}{OEL}{Ordered Epistemic Logic}
	\ouracronym{AFT}{AFT}{Approximation Fixpoint Theory}

	\makeatletter
	\def\ifenv#1{
	\def\@tempa{#1}%
	\def\@ttempa{#1*}%
	\ifx\@tempa\@currenvir
	\expandafter\@firstoftwo
	\else
	\expandafter\@secondoftwo
	\fi
	}
	\makeatother

	\newcommand{\ddrule}[4]{\ensuremath{#1 \leftarrow #2 & \{#3\} & #4}}
	\newcommand{\drule}[2]{\ensuremath{#1 & \leftarrow & #2}}

	\newcommand{\darule}[4]{\ensuremath{#1 \leftarrow #2 & \{#3\} & #4}}
	\newcommand{\arule}[2]{\ensuremath{#1 \, &\leftarrow \, #2}}

	\newcommand{\LNDRule}[2]{
	\ifenv{array}
	{\drule{#1}{#2}}
	{ \ifenv{align}
		{\arule{#1}{#2}}
		{\ifenv{align*}
		{\arule{#1}{#2}}
		{ERROR: using LDRule in unsupported environment: \@currenvir}
		}
	}
	}

	\newcommand{\LDRule}[4]{
	\ifenv{array}
	{\ddrule{#1}{#2}{#3}{#4}}
	{ \ifenv{align}
		{\darule{#1}{#2}{#3}{#4}}
		{\ifenv{align*}
		{\darule{#1}{#2}{#3}{#4}}
		{ERROR: using LDRule in unsupported environment: \@currenvir}
		}
	}
	}

	\NewDocumentCommand\LRule{m+g+g+g}{%
		\IfNoValueTF{#2}%
		{#1.&}{%
		\IfNoValueTF{#3}
		{\LNDRule{#1}{#2.}}
		{\LDRule{#1}{#2.}{#3}{#4}}%
		}
	}

	\NewDocumentCommand\CLRule{m+g}{%
	\ifenv{array}
	{\cdrule{#1}{#2}}
	{ \ifenv{align}
		{\carule{#1}{#2}}
		{\ifenv{align*}
			{\carule{#1}{#2}}
			{ERROR: using CLRule in unsupported environment: \@currenvir}
		}
	}
	}

	\NewDocumentCommand\carule{m+g}{%
		\IfNoValueTF{#2}
			{\ensuremath{#1.}}
			{\ensuremath{#1 \, &\cause \, #2}}}
	\NewDocumentCommand\cdrule{m+g}{%
		\IfNoValueTF{#2}
			{\ensuremath{#1.}}
			{\ensuremath{#1 & \cause & #2}}}

	\newcommand{\algrule}[4]{
	\hbox{{#1}:}& 
	\quad #2 ~\longrightarrow~ #3 
	\hbox{~ if } #4\\
	}

	\newcommand{\AlgoRule}[4]{
	\ifenv{array}
	{\algrule{#1}{#2}{#3}{#4}}
		{ERROR: using AlgoRule in unsupported environment: \@currenvir}
	}

	\newcommand{\ignore}[1]{}

	\newboolean{nocomments}
	\setboolean{nocomments}{false}

	\newboolean{commentmargin}
	\setboolean{commentmargin}{true}

	\newcommand{\namedcomment}[3]{%
		\ifthenelse{\boolean{nocomments}}%
		{}%IF no comments, write nothing
		{%Otherwise
			\ifthenelse{\boolean{commentmargin}}%
				{ {\color{#3} \marginpar{\color{#3}\sc #2}#1}  }%Name in margin
				{  {\color{#3} {\sc #2}: #1}  }%Name not in margin
		}%
	}
	\newcommand{\mnamedcomment}[3]{\ifthenelse{\boolean{nocomments}}{}{{\marginpar{ \color{#3}{\sc #2}:#1}}}}

	\newcommand{\bart}[1]{\namedcomment{#1}{bb}{red}}

	\usepackage{soul}

\usepackage[normalem]{ulem} % wavy underlines
\makeatletter
\font\uwavefont=lasyb10 scaled 700
\def\spelling{\bgroup\markoverwith{\lower3.5\p@\hbox{\uwavefont\textcolor{Red}{\char58}}}\ULon}
\def\grammar{\bgroup\markoverwith{\lower3.5\p@\hbox{\uwavefont\textcolor{LimeGreen}{\char58}}}\ULon}
\def\phrasing{\bgroup\markoverwith{\lower3.5\p@\hbox{\uwavefont\textcolor{RoyalBlue}{\char58}}}\ULon}

\newcommand\remove{\bgroup\markoverwith{\textcolor{red}{\rule[0.5ex]{2pt}{0.4pt}}}\ULon}
\makeatother

\usepackage{etoolbox}

\newcommand\setcitation[2]{%
  \csdef{mycommoncitation#1}{#2}}
\newcommand\getcitation[1]{%
  \csuse{mycommoncitation#1}}

\setcitation{IDP}{WarrenBook/DeCatBBD14}
\setcitation{idp}{WarrenBook/DeCatBBD14}
\setcitation{fodot}{tocl/DeneckerT08}
\setcitation{foid}{tocl/DeneckerT08}
\setcitation{FOID}{tocl/DeneckerT08}
\setcitation{cplogic}{journal/tplp/VennekensDB10}
\setcitation{CPlogic}{journal/tplp/VennekensDB10}
\setcitation{CPLogic}{journal/tplp/VennekensDB10}
\setcitation{CP}{fai/Rossi06}
\setcitation{cp}{fai/Rossi06}
\setcitation{EZCSP}{lpnmr/Balduccini11}
\setcitation{KR}{Baral:2003}
\setcitation{ASPComp2}{lpnmr/DeneckerVBGT09}
\setcitation{ASPComp3}{journals/tplp/CalimeriIR14}
\setcitation{ASPComp4}{conf/lpnmr/AlvianoCCDDIKKOPPRRSSSWX13}
\setcitation{ASPComp5}{journals/ai/CalimeriGMR16}
\setcitation{ASPComp6}{jair/GebserMR17}
\setcitation{ASPComp7}{tplp/GebserMR20}
\setcitation{CPSupport}{ictai/DeCat13}
\setcitation{CPsupport}{ictai/DeCat13}
\setcitation{functionDetection}{iclp/DeCatB13}
\setcitation{FunctionDetection}{iclp/DeCatB13}
\setcitation{fodot2asp}{corr/DeneckerLTV19} %TODO replace by journal publication if one is publishedr
\setcitation{Tarskian}{corr/DeneckerLTV19} %Same as the one above 
\setcitation{TarskianSemanticsASP}{corr/DeneckerLTV19} %Same as the one above 
\setcitation{Inca}{iclp/DrescherW12}
\setcitation{csp2asp}{ijcai/DrescherW11}
\setcitation{DPLLT}{cav/GanzingerHNOT04}
\setcitation{AspInPractice}{synthesis/2012Gebser}
\setcitation{ASPInPractice}{synthesis/2012Gebser}
\setcitation{clasp}{ai/GebserKS12}
\setcitation{oclingo}{kr/GebserGKOSS12}
\setcitation{clingo}{iclp/GebserKKOSW16}
\setcitation{gringo}{lpnmr/GebserST07}
\setcitation{cmodels}{aaai/GiunchigliaLM04}
\setcitation{inputster}{tplp/Jansen13}
\setcitation{DLV}{tocl/LeonePFEGPS06}
\setcitation{LearningPaper}{TPLP/BruynoogheBBDDJLRDV} %TODO replace by published version
\setcitation{clog}{iclp/BogaertsVDV14} %TODO replace by journal publication if one is published
\setcitation{foc}{iclp/BogaertsVDV14}
\setcitation{FOC}{iclp/BogaertsVDV14}
\setcitation{inferenceClog}{ecai/BogaertsVDV14}
\setcitation{examplesClog}{nmr/BogaertsVDV14b} %TODO replace by better publication if possible
\setcitation{AFT}{DeneckerMT00}
\setcitation{KBS}{iclp/DeneckerV08}
\setcitation{KBS-invitedtalk}{jelia/Denecker16}
\setcitation{KBPE}{inap/DePooterWD11}
\setcitation{lazyGrounding}{jair/CatDBS15} 
\setcitation{LazyGrounding}{jair/CatDBS15} 
\setcitation{lazygrounding}{jair/CatDBS15} 
\setcitation{lazygroundingASP}{ijcai/BogaertsW18} 
\setcitation{justifications}{lpnmr/DeneckerBS15} 
\setcitation{justificationsAlpha}{ijcai/BogaertsW18} 
\setcitation{ASP}{marek99stable}
\setcitation{satid}{sat/MarienWDB08}
\setcitation{lazyclausegeneration}{constraints/OhrimenkoSC09}
\setcitation{FP}{ACMCS/Hudak89}
\setcitation{GroundingWithBounds}{jair/WittocxMD10}
\setcitation{GroundWithBounds}{jair/WittocxMD10}
\setcitation{SAT}{faia/SilvaLM09}
\setcitation{HandbookOfSAT}{faia/2009-185}
\setcitation{LTC}{iclp/Bogaerts14}
\setcitation{SPSAT}{ictai/DevriendtBMDD12}
\setcitation{BreakID}{sat/DevriendtBBD16}
\setcitation{breakid}{sat/DevriendtBBD16}
\setcitation{LCG}{stuckeyLCG}
\setcitation{MiniZinc}{conf/cp/NethercoteSBBDT07}
\setcitation{minizinc}{conf/cp/NethercoteSBBDT07}
\setcitation{amadini}{cpaior/AmadiniGM13}
\setcitation{bootstrapping}{ngc/BogaertsJDJBD16}
\setcitation{Bootstrapping}{ngc/BogaertsJDJBD16}
\setcitation{GroundedFixpoints}{ai/BogaertsVD15}
\setcitation{PartialGroundedFixpoints}{ijcai/BogaertsVD15}
\setcitation{LogicBlox}{datalog/GreenAK12}
\setcitation{proB}{journals/sttt/LeuschelB08}
\setcitation{NaturalInductions}{KR/DeneckerV14} %TODO replace by journal publication if one is published
\setcitation{LP}{jacm/EmdenK76}
\setcitation{SMT}{faia/BarrettSST09}
\setcitation{AF}{ai/Dung95}
\setcitation{ADF}{kr/BrewkaW10}
\setcitation{af}{ai/Dung95}
\setcitation{adf}{kr/BrewkaW10}
\setcitation{ADFRevisited}{ijcai/BrewkaSEWW13}
\setcitation{adfrevisited}{ijcai/BrewkaSEWW13}
\setcitation{DefaultLogic}{ai/Reiter80}
\setcitation{DL}{ai/Reiter80}
\setcitation{AEL}{mo85}
\setcitation{minisat}{sat/EenS03}
\setcitation{completion}{adbt/Clark78}
\setcitation{ClarkCompletion}{adbt/Clark78}
\setcitation{wasp}{lpnmr/AlvianoDFLR13}
\setcitation{minisatid}{ictai/DeCat13}
\setcitation{lcg}{stuckeyLCG}
\setcitation{CEGAR}{jacm/ClarkeGJLV03}
\setcitation{cegar}{jacm/ClarkeGJLV03}
\setcitation{CuttingPlane}{or/DantzigFJ54}
\setcitation{kodkod}{tacas/TorlakJ07}
\setcitation{cdcl}{Marques-SilvaS99}
\setcitation{CDCL}{Marques-SilvaS99}
\setcitation{1UIP}{iccad/ZhangMMM01}
\setcitation{relevance}{ijcai/JansenBDJD16}
\setcitation{relevance-implementation}{aspocp/JansenBDJD16}
\setcitation{WFS}{GelderRS91}
\setcitation{wfs}{GelderRS91}
\setcitation{UnfoundedSet}{GelderRS91}
\setcitation{UFS}{GelderRS91}
\setcitation{stablesemantics}{iclp/GelfondL88}
\setcitation{StableSemantics}{iclp/GelfondL88}
\setcitation{shatter}{Shatter}
\setcitation{sbass}{drtiwa11a}
\setcitation{lparsemanual}{url:lparsemanual}
\setcitation{AIC}{ppdp/FlescaGZ04}
\setcitation{templates}{tplp/DassevilleHJD15}
\setcitation{templates2}{iclp/DassevilleHBJD16}
\setcitation{sat-to-sat}{aaai/JanhunenTT16}
\setcitation{sat-to-sat-qbf}{bnp/BogaertsJT16}
\setcitation{sat-to-sat-QBF}{bnp/BogaertsJT16}
\setcitation{sat-to-sat-SO}{kr/BogaertsJT16}
\setcitation{XSB}{SwiW12}
\setcitation{KCmap}{jair/DarwicheM02}
\setcitation{TLA}{DBLP:books/aw/Lamport2002}
\setcitation{EventB}{BookAbrial2010}
\setcitation{MX}{MitchellT05}
\setcitation{MIP}{Sierksma96}
\setcitation{perefectmodel}{minker88/Przymusinski88}
\setcitation{PerfectModel}{minker88/Przymusinski88}
\setcitation{perfectmodel}{minker88/Przymusinski88}
\setcitation{SafeInductions}{ijcai/BogaertsVD17}
\setcitation{AIC}{ppdp/FlescaGZ04}
\setcitation{aic}{ppdp/FlescaGZ04}
\setcitation{alpha}{lpnmr/Weinzierl17}
\setcitation{omiga}{jelia/Dao-TranEFWW12}
\setcitation{gasp}{fuin/PaluDPR09}
\setcitation{asperix}{lpnmr/LefevreN09a}
\setcitation{CTL}{lop/ClarkeE81}
\setcitation{AFT-AIC}{ai/BogaertsC18}
\setcitation{UltimateApproximator}{DeneckerMT04}
\setcitation{KripkeKleene}{Fitting85}
\setcitation{AFT-HO}{corr/CharalambidisRS18} %TODO replace
\setcitation{HereThere}{Heyting30}
\setcitation{dAEL}{ijcai/HertumCBD16}
\setcitation{SDD}{ijcai/Darwiche11}
\setcitation{HEX}{ijcai/EiterIST05}
\setcitation{wADF}{aaai/BrewkaSWW18}
\setcitation{wADFfix}{corr/BrewkaSWW18}
\setcitation{TransitionSystems}{jacm/NieuwenhuisOT06}
\setcitation{galliwasp}{lopstr/MarpleG12}
\setcitation{GalliWasp}{lopstr/MarpleG12}
\setcitation{clingcon}{tplp/BanbaraKOS17}
\setcitation{lp2sat}{birthday/JanhunenN11}
\setcitation{lp2mip}{LIU12}
\setcitation{lp2diff}{lpnmr/JanhunenNS09}
\setcitation{lp2acyc}{ecai/GebserJR14}
\setcitation{PB}{faia/RousselM09}
\setcitation{pb}{faia/RousselM09}
\setcitation{CuttingPlanes}{dam/CookCT87}
\setcitation{RoundingSAT}{ijcai/ElffersN18}
\setcitation{PRS}{aaai/DixonG02}
\setcitation{sat4j}{jsat/BerreP10}
\setcitation{SAT4J}{jsat/BerreP10}
\setcitation{mingo}{LIU12}
\setcitation{pbmodels}{lpnmr/LiuT05}
\setcitation{HEF-LP}{lpnmr/GebserLL07}
\setcitation{HCF-LP}{amai/Ben-EliyahuD94}
\setcitation{aspcore2}{tplp/CalimeriFGIKKLM20}
\setcitation{AspCore2}{tplp/CalimeriFGIKKLM20}
\setcitation{SemanticWeb}{book/AntoniouGHH12}
\setcitation{maxsat}{faia/LiM09}
\setcitation{MaxSAT}{faia/LiM09}
\setcitation{MAXSAT}{faia/LiM09}
\setcitation{}{}
\setcitation{}{}
\setcitation{}{}
\setcitation{}{}
\setcitation{}{}
\setcitation{}{}
\setcitation{}{}
\setcitation{}{}
\setcitation{}{}
\setcitation{}{}
\setcitation{}{}
\setcitation{}{}
\setcitation{}{}
\setcitation{}{}
\setcitation{}{}
\setcitation{}{}
\setcitation{}{}
\setcitation{}{}
  
\newcommand\refto[1]{%
      \ifcsname mycommoncitation#1\endcsname%
      \getcitation{#1}%
      \else%
      #1%
      \fi%
      }
      
\newcommand\mycite[1]{%
      \ifcsname mycommoncitation#1\endcsname%
   \cite{\getcitation{#1}}%
  \else%
    \cite{#1}%
  \fi%
}	
  
\usepackage{amsthm}
\usepackage{thmtools}

\declaretheorem[style=plain,	name=Theorem,		numberwithin=section]{thm}
\declaretheorem[style=plain,	name=Theorem,		numberlike=thm]{theorem}

\declaretheorem[style=plain,	name=Lemma,		numbered=no]   {lem*}

\declaretheorem[style=definition,	name=Definition,	numberlike=thm]{definition}

\declaretheorem[style=definition,	qed=$\blacktriangle$,	numberlike=thm]{example}

\declaretheorem[style=definition,	qed=$\blacktriangle$,	numbered=no]{ex*}

\declaretheorem[style=remark,	name=Notation,		numbered=no]{nota*}

\declaretheorem[style=remark,	qed=$\blacktriangle$,	name=Note,		numbered=no]{note*}

\ignore{

}
\usepackage{tikz}

\usepackage{verbatim}
\usetikzlibrary{arrows.meta,%
                petri,%
                topaths}%
\newcommand\defin{\m{\mathit{Def}}}
\renewcommand\theory{\m{\mathit{T}}}

\newcommand\citet[1]{{\color{red}AUTHORS}~\cite{#1}\xspace}

\newcommand\shacl{\textsc{shacl}\xspace}

\usepackage[capitalize]{cleveref}

\begin{document}

\maketitle

\begin{abstract}
\shacl is a W3C-proposed language for expressing structural
constraints on RDF graphs. The recommendation only specifies
semantics for \emph{non-recursive} \shacl; recently, some efforts 
have been made to allow \emph{recursive} \shacl schemas. 
In this paper, we argue that for defining and studying semantics of recursive \shacl, lessons can be learned from years of research in non-monotonic reasoning. 
We show that from a \shacl schema, a three-valued semantic operator can directly be obtained.
Building on \emph{Approximation Fixpoint Theory} (AFT), this operator immediately induces a wide variety of semantics, including a supported, stable, and well-founded semantics, related in the expected ways. 
By building on AFT, a rich body of theoretical results becomes directly available for \shacl. 
As such, the main contribution of this short paper is providing theoretical foundations for the study of recursive \shacl, which can later enable an informed decision for an extension of the W3C recommendation.
\end{abstract}
% 
% \section{Introduction}

\section{Introduction}

The \emph{Semantic Web} \mycite{SemanticWeb} extends the World-Wide Web with machine-interpretable data. 
The de facto standard of this web is that data is stored and published in the 
 Resource Description Framework \cite{rdf11primer},
i.e., as a set of triples, referred to as a \emph{graph}, 
often extended with semantic information expressed in OWL \cite{owldl}. 
While in principle the Semantic Web is open and every agent can represent their data however they want, from the perspective of applications, or when consuming RDF data in some other way, it can often be useful to know which structural properties an RDF graph in question satisfies. In other words, there is a need for a declarative language for describing \emph{integrity constraints} on RDF graphs. 

Several proposals have emerged to fill this need, the most prominent of them being ShEx \cite{shex}
and \shacl \cite{shacl}. 
Both approaches start from the notion of a \emph{shape}: a structural property that a node in an RDF graph can satisfy, e.g., the shape of ``things that have a name'' will be satisfied for those nodes that appear as the subject in a triple with predicate \texttt{foaf:name}. 
On top of a language for defining such shapes, the two proposals also have a mechanism for \emph{targeting}: for specifying which nodes \emph{should} satisfy which shapes, e.g., to declare that ``All persons should satisfy the has-a-name shape''.

In our paper, we will build on a formalization of \shacl
\cite{corman}, which has revealed a striking similarity between 
shapes and concepts descriptions, as known from description logics
\cite{dlintro}; 
we recently deepened this connection further \cite{ruleml/BogaertsJV21}. 
Following this line of research, we formalize a \shacl schema as a tuple $(\defin,\theory)$, where $\defin$ is a set of rules of
the form 
% \begin{align*} 
$
 s \lrule \concept
$ 
% \end{align*}
 and $\theory$ is a set of concept inclusions of the form 
% \begin{align*}
$
 \concept \subseteq s
 $
% \end{align*}
with  $  \concept$ a shape (i.e., a concept description) and $s$ a shape name. 
$\defin$ defines the shapes in terms of relations in the RDF graph and \theory contains the so-called \emph{targeting} constraints. % \footnote{In actual \shacl, the types of concept descriptions allowed in targeting constraints are limited.}

The W3C recommendation for \shacl only specifies the semantics when $\defin$ is \emph{non-recursive}, i.e., if no shape is defined in terms of itself, but recently, some efforts have been made to lift this restriction, i.e., to define a supported and a stable semantics for \emph{recursive} \shacl \cite{corman,andresel}. 
% In this direction, 
% \citet{corman} have defined a supported semantics and \citet{andresel} presented a \emph{stable} semantics, making use of a level mapping, inspired by the one of \citet{tplp/HitzlerW05}. % to define the semantics. 

In this paper, we put forward another principled way to define semantics of recursive \shacl, building on \emph{Approximation Fixpoint Theory} (AFT), an abstract lattice-theoretic framework originally designed to unify semantics of non-monotonic logics \mycite{AFT} with applications, among others, in (extensions of) logic programming, autoepistemic logic, default logic, abstract argumentation, and active integrity constraints  \cite{DeneckerMT03,tplp/PelovDB07,journals/ai/Strass13,ai/LiuBCYF16,tplp/CharalambidisRS18,ai/BogaertsC18}. 
There are several advantages to defining semantics of \shacl in this way:
\begin{compactitem}
\item It is \emph{simple} and \emph{straightforward}: 
the power of AFT, comes largely from the fact that all that is required to apply it, is to define a (three-valued) semantic operator (similar to Fitting's immediate consequence operator for logic programs \cite{tcs/Fitting02}).
In many domains (including \shacl), there is a natural choice for such an operator; AFT then immediately induces all major classes of semantics. %; \shacl is no exception in this respect. 
\item It provides \emph{confidence}: AFT guarantees that the developed semantics follow well-established principles in nonmonotonic reasoning. %, rather than ad-hoc solutions in which one can encounter issues that have been solved many times before. %, for instance that stable semantics respect principles such as \emph{groundedness} \cite{ai/BogaertsVD15}. 
% All the application-specific work that needs to be done is to define a suitable lattice of ``interpretations'', and a suitable three-valued operator on this lattice.  
% Defining such an operator is often significantly easier than directly defining a semantics. 
Even in case semantics are already defined, applying AFT can be a \emph{sanity check}. A striking example of this is the fact that applications of AFT uncovered some issues in the semantics of (weighted and non-weighted) Abstract Dialectic Frameworks \cite{journals/ai/Strass13,\refto{ADFRevisited},aaai/Bogaerts19}. 
\item It provides access to a \emph{large body of theoretical results}, including theorems on \emph{stratification} \cite{tocl/VennekensGD06,tocl/BogaertsC21}, \emph{predicate introduction} \cite{VennekensMWD07}, and \emph{strong equivalence} \cite{amai/Truszczynski06}, thereby eliminating the need to  ``reinvent the wheel'' by rediscovering these results in each of the separate domains.
\end{compactitem}
In a nutshell, our main contribution is establishing formal
foundations for the study of recursive \shacl.

\renewcommand\voc{\m{{\color{red}\Sigma}}}
\renewcommand\Sigma{\m{{\color{red}VOC}}}

\section{Formalization of \shacl}
%  \begin{remark}
In \emph{actual \shacl},  semantics is defined in terms of RDF graphs, but we recently showed how to reduce this to the logical setting \cite{ruleml/BogaertsJV21}. 
As such, for the purpose of the current paper, we focus on the logical setting and take abstraction of RDF graphs. 
Throughout this paper, we (implicitly) fix a \emph{vocabulary}, i.e., a set of \emph{node names} (denoted $N$), \emph{property names} (denoted $P$), and \emph{shape names} (denoted $S$). 
% 
% 
% 
% not interpretations, but RDF graphs are said to validate a schema. We showed precisely  how to reduce that case to the logical setting;  building on that work, we do not consider graphs explicitly here.  
% %  \end{remark}
% 
%  
% 
% \todo{Can we make $\Sigma$ implicit in the rest of the paper?
% The repeated occurrences of "\Sigma" throughout the paper makes the notation artificially heavy (in particular the repetition of "Sigma \cap").
% It is also used inconsistently ("\Sigma-interpretation" vs "interpretation over \Sigma").
% I wonder if it could be said in Section 2 that \Sigma is implicit in the rest of the paper.
% Or maybe use N_\Sigma (with subscript) as a shortcut for "N \cap Sigma", etc. (so that "Sigma \cap (N \cup P)" becomes "N_\Sigma \cup P_\Sigma").
% }
% 
% We define the formal semantics of \shacl, following our earlier work \cite{ruleml/BogaertsJV21}. 
% We assume three disjoint, infinite universes $N$, 
% $P$, and $S$ of 
% respectively.\footnote{In practice, node, property, and shape names
% are IRIs \cite{rdf11primer}, so the disjointness assumption would
% not hold.  This assumption is only made to simplify notation. %simplicity
% % of notation; it can be avoided if we make the notation for
% % vocabularies and interpretations more complicated.
% }
A node name $c$ is also referred to as a \emph{constant}, a property name $p$ as a \emph{binary predicate symbol} and a shape name $s$ as a \emph{unary predicate symbol}. 
% 
% A vocabulary $\voc$ is a subset of $N\cup P\cup S$. 
\emph{Path expressions} $E$ and \emph{shapes} $\phi$  are defined as follows: 
% by the following grammar:
\begin{align*}
  E & ::=  p \mid p^- \mid E \cup E \mid E \circ E \mid E^* \mid E?\\
  \concept &::= \top \mid s \mid \{c\} \mid \concept \land \concept \mid
  \concept \lor \concept \mid \neg \concept \mid \forall E.\concept \mid 
  {}\geqn{n}{E}{\concept} \mid  \eq(E,E)\mid \disj(E,E)\mid \closed(Q)
\end{align*}
where $p$, $c$, and $s$ represent property names, node names, and shape names respectively, 
and $n$ stands for non-zero natural numbers.
When developing our semantics, we will treat $\phi_1\lor\phi_2$ as an abbreviation of $\lnot(\lnot \phi_1\land\lnot \phi_2)$ and $\forall E.\phi$ as an abbreviation for $\lnot \geqn{1}{E}{\lnot \phi}$. The reason for having them in the syntax is to enable the comparison with existing semantics in \cref{sec:comparison}.
% A node name $c$ is also referred to as a
% \emph{constant}.

% \todo{Drop disjunction and universal quantification? Only mention when needed?}

% Shapes are evaluated in
% \emph{interpretations}. 
As usual, an interpretation $I$ consists of
\begin{inparaenum}
\item
a set $\dom^I$, called the \emph{domain} of
$I$;
\item
for each constant $c $, an element $\semi c \in
\dom^I$; 
\item for each shape name $s$, a set $\semi s\subseteq \dom^I$; and
\item
for each property name $p$, a binary relation $\semi
p$ on $\dom^I$.
\end{inparaenum}
When we say a \emph{graph}-interpretation, we mean an interpretation that only consists of a domain and an interpretation for node and property names (not shape names). Intuitively, such an interpretation corresponds to an RDF graph.
A path expression $E$ evaluates in $I$ to a binary relation $\semi E$ on $\dom^I$,
and a shape $\phi$ to a subset $\semi \phi$ of
$\dom^I$, as defined in Tables \ref{tab:sempath} and \ref{tab:sem}.

% \todo{Say $\concept$ does not mention shape names? And clarify ``actual shacl''}

 \begin{table}[t]
 \small
\parbox{.4\linewidth}{
\centering
 \begin{tabular}{@{}ll@{}}
  \toprule
   $E$ & $\iexpr{E}{I}$ \\
   \midrule
   $p^-$ & $\{(a,b) \mid (b,a) \in \iexpr{p}{I}\}$\\
   $\iexpr{E_1 \cup E_2}{I}$ & $\iexpr{E_1}{I} \cup \iexpr{E_2}{I}$\\
   $\iexpr{E_1 \circ E_2}{I}$ & $\{(a,b) \mid \exists c: (a,c) \in \iexpr{E_1}{I} $\\&\quad$ \land (c,b) \in \iexpr{E_2}{I} \}$\hspace{-10pt} \\
  $\iexpr{E^*}{I}$ &the reflexive-transitive\\&\quad closure of
  $\iexpr{E}{I}$\\
  $\iexpr{E?}{I}$ &  $\iexpr{E}{I} \cup \{(a,a)\mid a\in\dom^I\}$\\
  \bottomrule
  \end{tabular}
  %BART: IJCAI REQUIRES CAPTIONS UNDER TABLES 
  \caption{Semantics of a path expression $E$ in an interpretation $I$.} \label{tab:sempath}
}
\hfill
\parbox{.57\linewidth}{
\centering
\begin{tabular}{@{}ll@{}}
\toprule
  $\concept$ & $\iexpr{\concept}{I}$\\
  \midrule
  $  \top$ & $ \dom^I$ \\
  $\{c\}$ & $\{c^I\}$\\
  $\concept_1 \land \concept_2$ & $\iexpr{\concept_1}{I} \cap \iexpr{\concept_2}{I}$\\
%   $\concept_1 \lor \concept_2$ & $\iexpr{\concept_1}{I} \cup \iexpr{\concept_2}{I}$\\
  $\neg \concept_1$ & $\dom^I \setminus \iexpr{\concept_1}{I}$\\
  $\geqn{n}{E}{\concept_1} $ & $\{a \in \dom^I \mid \sharp (\semi\phi_1 \cap \iexpr{E}{I}(a)) \geq n\} $\\
  $\eq(E_1,E_2) $ & $\{a \in \dom^I \mid
\semi{E_1}(a) = \semi{E_2}(a)\}$ \\
%\begin{aligned}[t]
%& \forall b:(a,b) \in \iexpr{E_1}{I}\: \text{iff}\\[-0.75ex]
%& \qquad \qquad \qquad (a,b) \in \iexpr{E_2}{I}\}
%\end{aligned}$\\
  $\disj(E_1,E_2) $ & $\{a \in \dom^I \mid
\semi{E_1}(a) \cap \semi{E_2}(a) = \emptyset\}$ \\
%\begin{aligned}[t]
% \lnot \exists b: (a,b) \in \iexpr{E_1}{I} \land {} \\[-0.75ex]
% \qquad \qquad \qquad (a,b) \in \iexpr{E_2}{I}\}
%end{aligned}$\\
%   $\exists E.\self $ & $\{a \mid (a, a) \in \iexpr{E}{I}\}$\\
$\closed(Q) $ & $\{a \mid \semi{p}(a)=\emptyset \text{  for every } p\in P\setminus Q\} $\\
  \bottomrule
  \end{tabular}
  %BART: IJCAI REQUIRES CAPTIONS UNDER TABLES 
  \caption{Semantics of a shape $\concept$ in an interpretation $I$. 
  We use $\sharp X$ to denote the cardinality of $X$. For a binary
  relation $R$ and an element $a$, we use $R(a)$ to denote  $\{b \mid (a,b) \in R\}$.}\label{tab:sem}
}\vspace{-15pt}
\end{table}
 
 \noindent
  A \emph{\shacl schema} is a tuple $(\defin,\theory)$ with 
 \begin{compactitem}\item $\defin$ a set of rules of the form $s\lrule \concept$ with $s$ a shape name (referred to as the \emph{head} of the rule) and $\concept$ a shape (the \emph{body} of the rule), such that 
 each $s\in S$ appears exactly once in the head of a rule, and
  \item \theory a set of (concept) inclusions of the form $\concept \subseteq s$, with $\concept$ a shape that does not mention any shape names\footnote{The condition that $\phi$ does not mention any shape names will entail that the target query can be evaluated on the underlying graph, without knowledge of the shape definitions. In actual SHACL, the condition is even more limited; there only very specific queries are allowed as targets.}, and $s$ a shape name.
 \end{compactitem}
A shape name $s_1$ \emph{depends} on shape name $s_2$ in \defin if there is a rule $s_1\lrule \concept$ in \defin, and $s_2$ or some shape name that depends on $s_2$ occurs in $\concept$. A  schema is \emph{non-recursive} if no shape name depends on itself. 
 
 If $(\defin,\theory)$ is  a \emph{non-recursive} \shacl schema, and $I$ a graph-interpretation, then  $I$ can be uniquely extended to an interpretation $I'$ such that for each $s\in S$, $\iexpr{s}{I'} = \iexpr{\concept}{I'}$ if $s\lrule\concept\in\defin$. % and $\iexpr{s}{I'}=\emptyset$ if there is no such rule for $s$. 
 In that case, we say that $I$ \emph{validates} with respect to $(\defin,\theory)$ if $\iexpr{\phi}{I'}\subseteq \iexpr{s}{I'}$ for each inclusion $\phi\subseteq s$ in $\theory$. 
 For the \emph{recursive} case, the situation is less obvious; in \cref{sec:recursive}, we use approximation fixpoint theory to study the different semantic options that arise, but before doing so, we recall  preliminaries on AFT.

\section{Approximation Fixpoint Theory}
% \todo{Maybe "lattice" can be used throughout the paper, rather than "complete lattice" (or did I miss some usage of "lattice" in the paper that denotes either a complete or a semi-lattice?).
% }

% In this section we repeat the basics of approximation fixpoint theory and its application in logic programming.
A \emph{complete lattice} $\langle L,\leq\rangle$ is a set $L$ equipped with a partial order $\leq$, such that every 
set $S\subseteq L$ has a  least upper bound and a greatest lower bound. %, denoted $\lub(S)$ (or $\bigor S$) and $\glb(S)$ (or $\bigand S$) respectively.
A complete lattice has a least element $\bot$ and a greatest element $\top$. 
An operator $O:L\to L$ is \emph{monotone} if $x\leq y$ implies that $O(x)\leq O(y)$. An element $x\in L$ is  a \emph{fixpoint} of $O$ if $O(x)=x$. 
Every monotone operator $O$ in a %chain complete poset 
complete lattice has a least fixpoint, denoted $\lfp(O)$. %, which is also $O$'s least prefixpoint.
% \todo{define CPO}

% \todo{No usage is made in the paper of fact that $(L^2, \leqp)$ is a lattice (only fixed points over $L^c$ are considered).
% }

Given a lattice $L$, AFT uses a bilattice 
$L^2$.  We define \emph{projections} for pairs as usual:
$(x,y)_1=x$ and $(x,y)_2=y$.  Pairs $(x,y)\in L^2$ are used to
approximate elements in the interval $[x,y] = \{z\mid x\leq
z\wedge z\leq y\}$. We call $(x,y)\in L^2$ \emph{consistent} if $x\leq
y$, that is, if $[x,y]$ is non-empty. We use $L^c$ to denote the set
of consistent elements. 
% Elements $(x,x) \in L^c$ are called
% \emph{exact}. 
The \emph{precision
  order} on $L^2$ is defined as $(x,y) \leqp (u,v)$ if $x\leq u$
and $v\leq y$. If $(u,v)$ is consistent, this means that $(x,y)$
approximates all elements approximated by $(u,v)$.  
% If $L$ is a complete lattice, then so is
% $\langle L^2,\leqp\rangle$.

In its original form, AFT makes use of \emph{approximators}, which are operators on $L^2$, but \cite{DeneckerMT04} showed that all the consistent fixpoints studied in AFT are uniquely determined by an approximator's restriction to $L^c$ and developed a theory of \emph{consistent approximators}. 
% AFT studies fixpoints of lattice operators $O:L\ra L$ through operators approximating $O$.
 An operator $A: L^c\to L^c$ is a \emph{consistent approximator} of $O$ if it is \leqp-monotone and \emph{coincides with $O$ on $L$} (meaning $A(x,x) = (O(x),O(x))$ for all $x\in L$).
% Approximators are
% internal in $L^c$ (i.e., map $L^c$ into $L^c$).
% \citet{DeneckerMT04} showed that the consistent fixpoints of interest (supported, stable, well-founded) are uniquely
% determined by an approximator's restriction to $L^c$ and hence that it suffices to define what they called a \emph{consistent approximator} (a \leqp-monotonic operator $L^c\to L^c$ that coincides with $O$ on $L$); this is what we will do in this paperf. 
% 
AFT studies fixpoints of $O$ using fixpoints of $A$. 
\begin{compactitem}
% \item A \emph{(partial) $A$-supported fixpoint} is a fixpoint of $A$. 
\item The \emph{$A$-Kripke-Kleene fixpoint} is the $\leqp$-least fixpoint of $A$; it approximates all fixpoints of $O$. 
\item A \emph{partial $A$-stable fixpoint} is a pair  $(x,y)$ such that $x=\lfp(A(\cdot,y)_1)$ and $y=\lfp(A(x,\cdot)_2)$, where $A(\cdot,y)_1:L\to L$ maps $z$ to $A(z,y)_1$ and similarly for $A(x,\cdot)_2$. 
\item The \emph{$A$-well-founded fixpoint} is the least precise ($\leqp$-least) partial $A$-stable fixpoint.
\item An \emph{$A$-stable fixpoint} of $O$ is a fixpoint $x$ of $O$ such that $(x,x)$ is a partial $A$-stable fixpoint. 
\end{compactitem}

These definitions allow reconstructing all major %equally-named 
logic programming semantics by taking for $O$ Van Emden and Kowalski's immediate consequence operator $T_P$ \mycite{LP} and for $A$ Fitting's three- (or four-) valued extension $\Psi_P$ \cite{tcs/Fitting02}.

\section{Fixpoint Semantics for Recursive \shacl}
\label{sec:recursive}
For the rest of this paper, we fix a  \shacl schema $(\defin,\theory)$ and a graph-interpretation $I$. 
% Without loss of generality, we also assume that \emph{every shape in $\voc \cap S$ has a defining rule} (this can be obtained by adding a rule $s\lrule   \lnot \top$ to \defin). 
% 
We already mentioned that if \defin  is non-recursive, it uniquely induces a  complete interpretation $I'$ in which all constraints in \theory are to be verified. 
When \defin is recursive, however, the situation becomes more complex. 
On the one hand, there is a range of possible semantics dealing with recursion.
% \maxime{Deze laatste zin leest vreemd vind ik, misschien: ``On the one hand, there is a range of possible semantics dealing with recursion.''}
% AFT will be our guide in this zoo of different semantics.
On the other hand, some of the semantics yield not a single interpretation $I'$, but either a set of them, or a \emph{three-valued} interpretation. 
This will give us a choice between \emph{brave} and \emph{cautious} validation; the focus of this paper is on the treatment of negation, but we briefly discuss brave and cautious validation below. 
% For semantics resulting in a \emph{set} of models, this boils down to making a distinction between whether the constraints in  \theory hold in \emph{some} or in \emph{all} extensions of $I$; for three-valued semantics, the difference can be made between there not being any violations (each constraint in \theory is either true or unknown, but not false) or having the guarantee that the constraints in \theory are satisfied (each constraint being true). 
% The focus of this paper is on the first point, but the second dimension can also play a role when developing a standard for recursive \shacl. 
% \mtodo{ Defining brave and cautious would be clearer (and possibly more concise) that this half-informal description.
%  }\mbart{Tricky with a mix of two anhd threevalued semantics}

\newcommand\pinter{\m{\mathcal{I}}}
\newcommand{\oper}{\m{T_{\defin}}}
\newcommand{\app}{\m{\Psi_{\defin}}}

To apply AFT, the first step to take is to determine a suitable lattice. 
In our case, the obvious candidate is the lattice  $L_I$ (from now on, denoted  $L$) of all interpretations $I'$ with domain $\dom^I$ that agree with $I$ on $N\cup P$, or in other words, the set of interpretations that expand $I$. This set is equipped with the standard truth order, $I_1 \leqt I_2$ if $\iexpr{s}{I_1}\subseteq \iexpr{s}{I_2}$ for all $s\in S$.  
% The choice of this order already has implications on the resulting AFT semantics since several semantics (notably well-founded and stable semantics) aim to minimize truth in the given order. 
The role of the semantic operator is to update the value of the interpretation of the shapes. In analogy with logic programming, its definition is straightforward: it maps the interpretation $I'$ to $\oper(I')$ such that for each shape name $s$ with defining rule $s\lrule \concept$, we define
$\oper(I')(s) = \iexpr{\concept}{I'}$.

% 
% \mtodo{There is no earlier mention in the paper of what "semantic operator" means, let alone what "the" semantic operator refers to.
% Maybe write "the operator over L to be approximated", or write explicitly that this is the "O" mentioned in Section 3 (using the letter "T" or "O" in both places would also improve readability, as mentioned above).}
With the lattice $\langle L, \leqt\rangle$, elements of $L^c$ are pairs $\pinter=(\pinter_1,\pinter_2)$ of two interpretations with $\pinter_1\leqt \pinter_2$; such pairs correspond one-to-one to \emph{three-valued interpretations} that assign each $s\in S $ a function $\dom\to\{\ltrue,\lfalse,\lunkn\}$, mapping $a$ to \ltrue if $a$ in $\iexpr{s}{\pinter_1}$, to $\lfalse$ if $a\not\in \iexpr{s}{\pinter_2}$ and to $\lunkn$ otherwise (in other words,  $\pinter_1$ represents what is \emph{certainly true} and $\pinter_2$ what is \emph{possibly true}). From now on, we  simply refer to elements of $L^c$ as three-valued interpretations. 

\newcommand\tiexpr[2]{\m{\iexpr{#1}{#2}}}
We can evaluate a shape $\concept$ in a three-valued interpretation $\pinter$ with a straightforward extension of Kleene's truth tables, as also used in previous studies of recursive SHACL \cite{corman,andresel}; for completeness, this is included in \cref{tab:3valsem}. This table makes use of the truth order $\leqt$ on truth values defined as $\lfalse\leqt\lunkn\leqt\ltrue$, and the  negation on truth values defined as usual: $\lnot\ltrue=\lfalse; \lnot \lfalse = \ltrue; \lnot \lunkn=\lunkn$. 
\begin{table}[h] 
\centering
\parbox{.67\linewidth}{
\centering
\small
\begin{tabular}{@{}ll@{}}
\toprule
$\concept$ & $\tiexpr{\concept}{\pinter}(a)$\\
\midrule
    $\top$ & $\ltrue$ \\
    $\{c\}$ & $\ltrue$ if $a=c$; $\lfalse$ otherwise\\
    $s$ & $\tiexpr{s}{\pinter}(a)$ \\
    $\neg \phi_1$ & $\neg \tiexpr{\phi_1}{\pinter}(a)$\\
    $\phi_1 \land \phi_2$ & $\min_{\leqt} (\tiexpr{\phi_1}{\pinter}(a),  \tiexpr{\phi_2}{\pinter}(a))$\\
%     $\phi_1 \lor \phi_2$ & $\max_{\leqt} (\tiexpr{\phi_1}{\pinter}(a),  \tiexpr{\phi_2}{\pinter}(a))$\\
    $\geqn{n}{E}{\phi_1} $ & $
                             \begin{cases}
                               \ltrue & \text{if}\; \sharp\{b\in \iexpr{E}{\pinter}(a) \mid 
                                   \tiexpr{\phi_1}{\pinter}(b) = \ltrue \} \geq n, \\
                               \lfalse & \text{if}\;  \sharp\{b\in \iexpr{E}{\pinter}(a) \mid 
                                   \tiexpr{\phi_1}{\pinter}(b) \leqt \lunkn \} < n , \\  
                               \lunkn  & \text{otherwise}
                             \end{cases}$ \\
    $\eq(E_1,E_2) $ & \ltrue if $a\in \iexpr{\eq(E_1,E_2)}{\pinter}$; $\lfalse$ otherwise \\
    $\disj(E_1,E_2) $ & \ltrue if $a\in \iexpr{\disj(E_1,E_2)}{\pinter}$; $\lfalse$ otherwise \\
    $\closed(Q)$ & \ltrue if $\iexpr{p}{\pinter}(a)=\emptyset$ for all $p\in P\setminus Q$; $\lfalse$ otherwise\\
    \bottomrule
  \end{tabular} 
  \caption{Three-valued semantics of shapes.}
  \label{tab:3valsem}
  }
\parbox{.3\linewidth}{
  \centering
\begin{tikzpicture}
\tikzstyle{grnode}=[draw,circle,fill=white,minimum size=4pt,
                            inner sep=0pt]
\draw (0,0) node [grnode] (a) [label=left:$a$] {};
 \draw (1,1) node [grnode] (c) [label=above:$c$] {};
  \draw (2,0) node [grnode] (b) [label=right:$b$] {};
  
 \draw (0,2.5) node [grnode] (e) [label=left:$e$] {};
 \draw (1,3.5) node [grnode] (d) [label=above:$d$] {};
 \draw (2,2.5) node [grnode] (f) [label=right:$f$] {};
 \draw (3.5,1) node [grnode] (Pfizer) [label=above:{\tiny$\mathit{Pfizer}$}] {};
 \draw (3.5,3.5) node [grnode] (Cough) [label=above:{\tiny$\mathit{Cough}$}] {};
\draw [{Latex[length=3mm]}-{Latex[length=3mm]}] (a) -- (c) node [sloped,midway,above ] {\tiny{$\mathit{closeTo}$}};
\draw [{Latex[length=3mm]}-{Latex[length=3mm]}] (c) -- (b) node [sloped,midway,above ] {\tiny$\mathit{closeTo}$};
\draw [{Latex[length=3mm]}-{Latex[length=3mm]}] (a) -- (b) node [sloped,midway,below] {\tiny$\mathit{closeTo}$};

\draw [{Latex[length=3mm]}-{Latex[length=3mm]}] (f) -- (e) node [sloped,midway,below ] {\tiny$\mathit{closeTo}$};
\draw [{Latex[length=3mm]}-{Latex[length=3mm]}] (e) -- (d) node [sloped,midway,above ] {\tiny$\mathit{closeTo}$};
\draw [{Latex[length=3mm]}-{Latex[length=3mm]}] (d) -- (f) node [sloped,midway,above] {\tiny$\mathit{closeTo}$};

\draw [-{Latex[length=3mm]}] (c) -- (Pfizer) node [sloped,midway,above] {\tiny$\mathit{vaccinated}$};
\draw [-{Latex[length=3mm]}] (d) -- (Cough) node [sloped,midway,above] {\tiny$\mathit{hasSymptoms}$};

\end{tikzpicture}

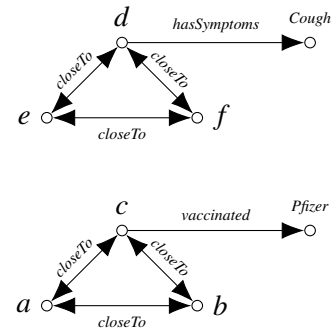
\captionof{figure}{Visual representation of the example interpretation.} 
\label{fig:ex}
  }

\end{table}

Once a three-valued evaluation of shapes is defined, an approximator is obtained directly: like the operator, the approximator updates the value of each shape symbol according to its defining rule: it maps \pinter to $\app(\pinter)$ where for each shape $s\in S$ defined by the rule $s\lrule \concept$, $\app(\pinter)(s) = \iexpr{\concept}{\pinter}$.

\begin{theorem}
 \app is a consistent approximator of $\oper$. 
\end{theorem}

At this point, AFT dictates what the supported models (fixpoints of \oper), (partial) stable models (\app-stable fixpoints), well-founded model (\app-well-founded fixpoint), and Kripke-Kleene model (\app-KK fixpoint) of \defin are. 
It is worth stressing that to arrive to this point, we made two choices. The first was our choice of order on the lattice. We opted here for the truth order, but its inverse would also have been a possible choice. 
Several of the semantics induced by AFT aim to minimize models in the chosen order for reasons of \emph{groundedness} \cite{ai/BogaertsVD15}, e.g., if $s$ has $s\lrule s$ as defining rule, in stable and well-founded semantics, our chosen order would result in no nodes satisfying $s$. % . For this reason, our choice of order boils down to imposing that ``a domain element only satisfies a given shape if the defining rule provides derives it'', which is a natural assumption here. 
The second choice we made is \emph{which three-valued truth evaluation to use}; we opted for the most obvious choice: a direct extension of Kleene's three-valued truth tables, which was used in other studies of recursive SHACL as well \cite{corman,andresel}. 
Given these choices, models of the different types (stable, well-founded, ... ) are defined by AFT, and hence semantics for brave and cautious validation under of each semantics are established. 
\begin{definition}
 Let $\sigma\in\{\mathit{KK},\mathit{WF}\}$ and let $\pinter$ be the $\sigma$-model. We say that 
$I$ \emph{cautiously (resp.\  bravely) $\sigma$-validates} with respect to $(\defin,\theory)$ if $\iexpr{\concept\land\lnot s}{\pinter}=\lfalse$ (resp.\  $\iexpr{\concept\land\lnot s}{\pinter}\leqt\lunkn$) for every $\concept\subseteq s$ in $\theory$. 

\noindent
Let $\sigma\in\{\mathit{St},\mathit{Sup}\}$ and let $M$ be the set of $\sigma$-models. We say that 
$I$ \emph{cautiously (resp.\  bravely) $\sigma$-validates} with respect to $(\defin,\theory)$ if 
$\iexpr{\concept\land\lnot s}{I'}=\lfalse$ for all (resp.\ for some) $I'\in M$ for every $\concept\subseteq s$ in $\theory$. 

\end{definition}

Let us illustrate the differences between the various types of models on a small example.%
\begin{example} \label{ex:first}
% \bart{Update voc}
 Consider binary predicates $\mathit{closeTo}$, $\mathit{hasSymptoms}$, and $\mathit{vaccinated}$ and an interpretation $I$ with domain $\{a,b,c,d,e,f,\mathit{Pfizer},\mathit{Cough}\}$, where $a,\dots,f$ represent people (divided in two cliques of three ``close'' friends); one person ($c$) is vaccinated and one person ($d$) shows Covid symptoms. This interpretation is visually  depicted in \cref{fig:ex}. 
% 
% 
% 
% 
% \begin{figure}
% 
% \end{figure}
% 
% 
%  
We define two shapes: the shape of people at risk (those who \begin{inparaenum}\item are not vaccinated and \item have symptoms or are close to someone at risk\end{inparaenum}) and the shape of people who can go to office (those who are not at risk), as formalized below:
\begin{align*}
  \small
    \mathit{atRisk} &\gets \lnot \geqn{1}{\mathit{vaccinated}}{\top} \land % \\&
     (\geqn{1}{ \mathit{hasSymptoms}}{\top} \lor \exists \mathit{closeTo}.\mathit{atRisk}) \\
    \mathit{canWork} &\gets \neg \mathit{atRisk} 
  \end{align*}
For this set of shape definitions, the unique stable model equals the well-founded model and states that $d$, $e$, and $f$ are at risk, while $a$, $b$, and $c$ can work. 
In the Kripke-Kleene model, $d$, $e$, and $f$ are again at risk, $c$ is not at risk (and hence can work), but for $a$ and $b$ it is unknown whether they are at risk. 
There are two supported models: the  stable model and one in which everyone except for $c$ is at risk. 
\ignore{
 Let $G$ be the graph consisting of three nodes:
  $\{\mathit{Tom}, \mathit{Ann}, \mathit{Bob}\}$ and which forms a
  clique through the $\mathit{closeTo}$ property. Furthermore, we add the tuple $(\mathit{Tom}, \mathit{yes})$ to the extension of $\mathit{hasSymptoms}$. The graph is depicted in Figure \ref{fig:example}.

  \begin{figure}
    \centering
    \includegraphics{graph-example.pdf}
    \caption{Example graph}
    \label{fig:example}
  \end{figure}

  Consider the following shapes:
  \begin{align*}
    \small
    \mathit{NeedsTest} & \gets \exists \mathit{hasSymptoms}.\mathit{yes} \lor \exists \mathit{closeTo}.\mathit{NeedsTest} \\
    \mathit{Healthy} & \gets \neg \mathit{NeedsTest} 
  \end{align*}
  There is one supported model $I=\{\mathit{NeedsTest}(\mathit{Tom}),$ $\mathit{NeedsTest}(\mathit{Ann}),$ $\mathit{NeedsTest}(\mathit{Bob})\}$. Next, consider the following shape:
  \begin{align*}
    \small
    \mathit{Safe} & \gets \exists \mathit{vaccinated}.\mathit{yes} \lor {}\leq_1 \mathit{closeTo}.\neg \mathit{Safe}
  \end{align*}
  Written in normal-form we obtain the set of shape definitions:
  \begin{align*}
    \small
    \mathit{Safe} & \gets \exists \mathit{vaccinated}.\mathit{yes} \\
     \mathit{Safe} & \gets \neg s_1 \\
     s_1 & \gets {}\geqn{2}{\mathit{closeTo}}{s_2} \\
     s_2 & \gets \neg\mathit{Safe}
  \end{align*}
  There are two possible 2-valued supported models: $\emptyset$ and
  $\{\mathit{Safe}(\mathit{Tom}),$ $\mathit{Safe}(\mathit{Ann}),$
  $\mathit{Safe}(\mathit{Bob})\}$. We refer to the latter model by
  $I$. There exists a proper level mapping for $I$. For all
  $b\in \{\mathit{Tom}, \mathit{Ann}, \mathit{Bob}\}$:
  $\lvl(\top, b) = 0$, $\lvl(\neg s_1, b) = 0$ and
  $\lvl(\mathit{Safe}, b) = 1$. Therefore, all nodes in $G$ are
  considered safe under $\lvl$-stable model semantics.

  However, this is not the case under $\twop{\defin, G}$-stable model
  semantics. $\twop{\defin, G}(\emptyset, I)_1 = \emptyset$, therefore
  $I$ is not an $\twop{\defin, G}$-stable model.

 FIVE INDIVIDUALS: CLIQUE OF THREE; VACCINATED GUY IN CONTACT WITH EVERYONE; Fifth person only in contact with the vaccinated guy. 
 
 HERE WE SAY: ``safe is: vaccinated or all friends are safe'' 
 This works well... 
 
 ALSO include something to verify. E.G. all citizens are safe.

 \bart{Maxime, can you write out this example, mabye with a tikz visualization of the input interpretation? ? }
 }
 \end{example}

% \todo{Define two- and three-valued semantic operator for SHACL. Should be straithgforward}
% 
% \bart{Maybe also mention that validation in three-valued/multi-model setting there is another dimension: skeptical vs vs credulous validation}
% 
% \maxime{Andresel calls it Brave vs Cautious reasoning. Corman uses
%   brave supported model sem, Andresel brave stable model sem. Andresel
%   discusses this a little.}
% \section{Properties}

% \todo{Comparison with translation-based 

% % \section{Properties of Semantics}
% \todo{Some interesting properties. 
% E.g. complexity results? 
% E.g. the fact that the semantics respects standard operations on bodies of rules, eg pushing negation, distributivity,... (should be easy)}

\section{Comparison with Existing Semantics}\label{sec:comparison}
% We are not the first to study semantics of recursive \shacl. 
Corman et al.~\cite{corman} already defined a supported semantics and Andre\c sel et al.\ \cite{andresel} a stable semantics for \shacl. For clarity, we refer to the existing semantics as CRS-supported and ACORSS-stable semantics, and to the semantics induced by AFT, e.g., as AFT-stable. Both of them focus on brave validation, but Andre\c sel et al.\ also mention the possibility of cautious validation.  The main results on correspondence between the semantics are summarized here. 

% \todo{MAXIME, can you remind me... how did they define it? Did they define a notion of supported model that ONLY looks at \defin, not at $T$? And then say something about validation? Or did they, in one shot go to validation? Might be important for deciding how to phrase the theorem}
% \maxime{Andresel: Ze definieren supported models, dan eisen ze dat er een lvl map bestaat, het valideert als het voldoet aan de targets. Ze hebben enkel ``constant targetting'' in beide papers. In een zekere zin doen ze validatie ``achteraf'': eerst de modellen en dan checken of de constanten op de juiste plaats zitten.}

In fact, 
Corman et al.\ \cite[Definition 5]{corman} already defined the three-valued immediate consequence operator \app (there denoted $\mathbf{T}$). While the focus of that work was on supported semantics, 
we now showed that in fact, by defining the approximator \app, they had everything at hand to define the full. 
Since Corman et al.\ also characterized supported models as fixpoints of \app (in their Definition 17), our semantics and theirs coincide.

% already defined the operator \oper and defined CRS-supported models as fixpoints of that operator 
% \bart{Corman en co hadden de DRIEWAARDIGE OPERATOR!}
\begin{theorem}
% An interpretation $I$ is a CRS-supported model of \defin if and only if it is an AFT-supported model. 
% 
$I$ CRS-validates $(\defin,\theory)$ if and only if $I$ bravely $\mathit{Sup}$-validates $(\defin,\theory)$. 
% \emph{some} AFT-supported model of \defin that expands $I$.
\end{theorem}

The situation is somewhat different for stable semantics, which Andre\c sel et al.\ \cite{andresel} defined in terms of level mappings. 

\newcommand\level{\m{\mathit{level}}}
\begin{definition}[\cite{andresel}]
 Let $I'$ be an interpretation. 
 A \emph{level assignment} for $I'$ is a function \level that maps tuples in $\{(\phi,a)\mid \iexpr{\phi}{I'}(a)=\ltrue\}$ to integers and satisfies 
 \begin{inparaenum}
  \item $\level(\phi_1\land\phi_2,a)= \mathrm{max}\{\level(\phi_1,a),\level(\phi_2,a)\}$,
  \item  $\level(\phi_1\lor\phi_2,a)= \mathrm{min}\{\level(\phi_i,a)\mid i\in \{1,2\}\land  \iexpr{\phi_i}{I'}(a)=\ltrue\}$,  
  \item $\level(\geqn{n}{E}{\concept})$ is the smallest $k\geq 0$ for which there are $n$ elements $b_1,\dots b_n \in\dom^{I'}$ such that $\level(\concept,b_i)\leq k$, $(a,b_i) \in\iexpr{E}{I'}$, and $\iexpr{\concept}{I'}(b_i)=\ltrue$, and \item $\level(\forall E. \concept, a) = \max(\{\level(\concept,b) \mid (a,b) \in \iexpr{E}{I'} \land \iexpr{\concept}{I'}{(b)}= \ltrue\}$.
 \end{inparaenum}
 
 A supported model $I'$ is an ARCOSS-stable model if there exists a level assignment for $I'$ such that $\level(s,a)>\level(\concept_s,a)$ for each rule $s\lrule\phi_s$ in $\defin$ and each $a$ with $\iexpr{s}{I'}(a) = \ltrue$.
\end{definition}

For  the correspondence in case of stable semantics, we recall a normal form of Andre\c sel et al.\ \cite{andresel}:
\begin{definition}
%   A set of shape definitions 
$\defin$ is in \emph{shape normal form} if all
  rules in $\defin$ have one of the
  following forms:
  \begin{center}
    \center
    \small
    \begin{tabular}[h]{lllll}
      $s\gets \top$ & $s\gets \{c\}$ & $s\gets \neg s'$ &
      $s\gets s' \land s''$ & $s\gets s' \lor s''$ \\ $s\gets \geqn{n}{E}{s'}$ &
      $s\gets \forall E.s'$ & $s\gets \eq(E, E')$ & $s\gets \disj(E, E')$ \\
    \end{tabular}
  \end{center}
\end{definition}

\begin{theorem}
%  Let \defin be a set of shape definitions over \voc, and $I'$ a \voc-interpretation. 
%  
 If $I'$ is an AFT-stable model of \defin, then it is also an ACORSS-stable model. 
 If \defin is in shape normal form, the converse also holds.
\end{theorem}
The difference between our stable semantics and the ACORSS-stable semantics is a \emph{semantic} (in terms of the standard three-valued truth evaluation) versus a \emph{syntactic} (the level mappings are defined in terms of the syntactic structure of the shapes) treatment of negation and is illustrated in the next example. 
% \todo{This sentence doesn't explain much (i.e. it does not add much to Theorem 5.3).}

\begin{example}[Example \ref{ex:first} continued]
Suppose that in our same interpretation, we wish to define a shape that identifies possible superspreaders. To do this, we say that a person is ``safe'' if they are vaccinated, or in contact with at most $1$ non-safe person. 
This can be formalized as: 
\begin{align*}
  \small
    \mathit{Safe} & \gets \exists \mathit{vaccinated}.\top \lor {}\leq_1 \mathit{closeTo}.\neg \mathit{Safe},
  \end{align*}
Where  $\leq_1$ is an abbreviation for $\lnot \geq_2$. 
With the interpretation of \cref{fig:ex}, there is a single AFT-stable model in which $a$, $b$, and $c$ are safe, but $d$, $e$, and $f$ are not. 
However, there are two ACROSS-stable models: the one mentioned above, and one in which everyone is safe, including the three-clique of non-vaccinated people. 
\end{example}

% \begin{example}
%   In case the \todo{condition from theorem} is not satisfied, the semantics do not coincide. 
%   \todo{Find natural sentence with double negation and recursion. Argue that in that case our semantics is the natural one. }
% \end{example}

\section{Conclusion}
% One line summary of our results.

When Corman et al.\ \cite{corman} defined the supported model semantics for \shacl, they showed how to translate \emph{actual \shacl expressions} (as specified by the W3C recommendation) into logical expressions in a language akin to description logics. For studying recursive \shacl expressions, they even already defined the operator $\app$. 
Hence, the only work left to obtain a rich family of semantics, was  observing that this operator is indeed an approximators and applying AFT. 
As such, we believe this paper establishes strong and formal foundations for the study of recursive \shacl. 
% In this paper, we focused on providing a formal foundation for the study of semantics of recursive \shacl. 
% The foundations are developed with remarkable ease after making two seemingly obvious choices, the whole family of semantics of recursive \shacl is immediately defined using AFT. 
Indeed, AFT does not just dictate how the semantics are to be defined, but immediately provides guarantees such as stratification and predicate introduction results that can be instrumental when developing concrete validators for recursive \shacl: we immediately obtain results about which transformations can safely be applied to our theories. 
% \bart{Werk van Corman meer ophemelen? Cfr email naar Jan} 

Our results have been presented without taking a stance on the choice of semantics. 
% this far without taking a stance on which semantics should be chosen. 
It is important to realize though, that if one wants to view \defin as an (inductive) \emph{definition} of the shapes, it has been argued repeatedly that the well-founded semantics correctly formalizes this \cite{KR/DeneckerV14}. 
Under the well-founded semantics, \shacl integrates first-order constraints (of a restricted form)  with inductive definitions and aggregates, and hence can be seen as a fragment of the language $\mathrm{FO}(\mathrm{ID},\mathrm{Agg})$, the formal foundation of the \idp language \mycite{IDP}. 
It is a topic for future work to investigate whether this can be exploited for either extending \shacl, or for developing alternative validation mechanisms, building on  \idp. 

% \todo{Other thoughts for future work? } 

% No claims made about what a ``good''

% \newpage
% 
% 

\noindent
\textbf{Acknowledgements }
This research was supported by the Flemish Government in the ``Onderzoeksprogramma Artifici\"ele Intelligentie (AI) Vlaanderen'' programme and by FWO Flanders project G0B2221N.

\bibliographystyle{eptcs}

%REDUCE BIB SIZE
\let\OLDthebibliography\thebibliography
\renewcommand\thebibliography[1]{
\OLDthebibliography{#1}
\setlength{\parskip}{0pt}                                                                                                                                                \setlength{\itemsep}{0pt}        
\scriptsize
}

\bibliography{idp-latex/krrlib,database-shortened,extra}
% \newpage
% \todo{This should be on page at most 8} 
\end{document}

%%% Local Variables:
%%% mode: latex
%%% TeX-master: t
%%% End: